\documentclass[prd,aps,floats,epsfig,twocolumn]{revtex4}
\usepackage{graphicx}
\newcommand{\beq}{\begin{equation}}
\newcommand{\eeq}{\end{equation}}
\newcommand{\bea}{\begin{eqnarray}}
\newcommand{\eea}{\end{eqnarray}}

\begin{document}

\title{When Did Cosmic Acceleration Start ?}
\author{Alessandro Melchiorri, Luca Pagano, Stefania Pandolfi}
\affiliation{Physics Department and Sezione INFN, 
University of Rome ``La Sapienza'', 
Ple Aldo Moro 2, 00185, Rome, Italy}

\begin{abstract}

A precise determination, and comparison, of the epoch of the onset of cosmic 
acceleration, at redshift $z_{acc}$, and of dark energy domination, at $z_{eq}$, provides an interesting measure with which to parameterize dark energy models.
By combining several cosmological datasets we place constraints
on the redshift and age of cosmological acceleration.
For a $\Lambda$CDM model, we find the constraint $z_{acc}=0.76\pm0.10$ 
at $95 \%$ c.l., occurring $6.7 \pm 0.4$ Gyrs ago. Allowing a constant 
equation of state but different from $-1$ 
changes the constraint to $z_{acc}=0.81\pm0.12$
($6.9 \pm 0.5$ Gyrs ago)
and  $z_{eq}=0.48\pm0.14$ ($4.9 \pm 0.9$ Gyrs ago)
, while dynamical
models markedly increase the error on the constraint  with 
$z_{acc}=0.81\pm0.30$ ($6.8 \pm 1.4$ Gyrs ago)
and $z_{eq}=0.44\pm0.20$ ($4.5 \pm 1.0$ Gyrs ago). 
Unified dark energy models as Silent Quartessence yield: 
$z_{acc}=0.80\pm0.16$ ($6.8\pm0.6$ Gyrs ago).

\end{abstract}
\maketitle

\section{Introduction}

The existence of a dark and unclustered energy component 
responsible for more than $70 \%$ of the overall density
of our universe is now suggested at high significance 
by most of the latest cosmological data 
(see e.g. ~\cite{spergel},  ~\cite{seljak}).

A cosmological constant provides a possible candidate for 
the dark energy component, but it needs to have its
initial conditions properly `tuned' in order to dominate 
the universe expansion at precisely the present time. 
Indeed the energy density in a 
cosmological constant $\rho_\Lambda$ does not evolve, 
while both matter ($\rho_m$)
and radiation ($\rho_r$) energy densities evolve 
rapidly with the expansion of the universe.
The small current value 
of $\rho_\Lambda$ implies an extreme fine-tuning of 
initial conditions with $\rho_\Lambda/\rho_r \simeq 10^{-123}$ 
at the Planck time (when the temperature of the universe was 
$T \sim 10^{19}$ GeV), or $\rho_\Lambda/\rho_r \simeq 10^{-55}$ 
at the time of the electroweak phase transition ($T \sim 100$
GeV). Moreover, $\rho_\Lambda$ lies in a very small window, 
since a slightly larger value makes the universe accelerate much 
before the present epoch, thereby inhibiting structure formation, 
while a negative value may cause the universe to
re-collapse.

The cosmological constant problem has, subsequently, motivated several 
``dynamical'' alternatives (see e.g. \cite{copeland07}
for a recent and very complete review) as a slowly-rolling scalar field,
``quintessence''~\cite{Wetterich:fm}-\cite{Caldwell:1997ii}, or
a ``k-essence''  scalar field with non-canonical kinetic terms in the 
Lagrangian ~\cite{Armendariz-Picon:1999rj},
string-inspired models such as the contribution of 
nonlinear short distance physics to vacuum
energy~\cite{transplanck}, and modified Friedman equations at late
time~\cite{cardassian} or large distances~\cite{dgp}.

Other possibilities include anthropic arguments 
(\cite{barrow86,vilenk00,wein01}) and  ``backreaction'' of non 
linear inhomogeneities (see \cite{kolb07}, but see
also \cite{Flanagan:2005dk,hirata}).

It is plausible that a solution to the dark energy problem
could be found by identifying a time correlation between the epoch of 
appearance of this exotic component and a well understood 
and physically motivated event such as the time of matter-radiation 
equality, the origin of non linear structures
or, ultimately, life. It is, therefore, clear 
that a first crucial measurement that has to be made
is the determination of the redshift and time of dark energy domination.
Evidence for dark energy at very high redshifts ($z>1$),
when the cosmological constant is negligible, would indeed favor 
models based on scalar fields, possibly coupled to dark matter \cite{Amendola:1999er, Bean:2000zm}. 
While the appearance of dark energy at lower redshifts ($z<0.2$) 
would, on the contrary hint  at ``phantom'' ($w < -1$) models.
Anthropic principle arguments are definitely 
less appealing if dark energy dominates well after the epoch of 
formation of terrestrial planets. At the same time, backreaction
models could be perceived as much less convincing if dark energy 
starts in a time when nonlinear structures are already 
well present and formed.

The starting point of cosmic 
acceleration, however, is not a model independent
quantity.  If the universe is in accelerated expansion today we can
identify two crucial epochs.  Firstly, the epoch of equality between matter and
the dark energy component, at redshift $z_{eq}$, defined as
\begin{equation}
\rho_m(z_{eq})=\rho_X(z_{eq})
\end{equation}
\noindent where $\rho_m(z)$ and $\rho_X(z)$ are the
energy densities of the matter and dark energy components
at redshift $z$ respectively. 
This epoch is generally different from,
and follows, the redshift $z_{acc}$ when the universe 
started to accelerate, defined as
\begin{equation}
q(z_{acc})={-{1 \over H^2}{\ddot a \over a}} (z_{acc})=0\\
\end{equation}
\noindent where $H=\dot a / a$ is the Hubble parameter at redshift 
$z$. The two epochs are model dependent and distinct.
In the simple case of a cosmological constant, for example,
the two redshifts are not equal and the following simple 
relation:
\begin{equation}
z_{acc}=2^{1/3}(1+z_{eq})-1
\end{equation}
\noindent holds. The age of the universe at each of those redshifts 
can then be easily computed from:
\begin{equation}
t(z)=\int_z^{\infty} {dz \over {(1+z)H(z)}}
\end{equation}
\noindent and compared with the current age of the universe $t_0$.

It is clear that constraints on the model-dependent quantities 
$z_{acc}$, $z_{eq}$, $t(z_{eq})$ and $t(z_{acc})$ can 
provide relevant information for several
studies. In this paper, we focus on constraining these
quantities with current cosmological data with the goal
of clarifying the following points: how model independent are the
constraints on the epoch of dark energy domination ? how does a different choice
of cosmological datasets or parameters affect those constraints ? 
Finally, are the constraints, derived in a general dark energy
scenario, consistent with the predictions of a cosmological constant ?

Our paper is organized as follows, in the next section we 
introduce our data analysis method, describing the 
datasets and dark energy parameterizations adopted.
In section III we  present the results of our analysis 
and in section IV we derive our conclusions.

\section{Likelihood analysis}
\label{sec3}

The method we adopt is based on the publicly available Markov Chain Monte Carlo
package \texttt{cosmomc} \cite{Lewis:2002ah}. We sample the following
dimensional set of cosmological parameters, adopting flat priors on them:
the physical baryon and CDM densities, $\omega_b=\Omega_bh^2$ and
$\omega_c=\Omega_ch^2$, the ratio of the sound horizon to the angular diameter
distance at decoupling, $\theta_s$, the scalar spectral index, $n_{s}$,
and the optical depth to reionization, $\tau$.  
Furthermore, we consider purely adiabatic
initial conditions and the possibility of curved universes,
$\Omega_{tot}\neq1$.
We also consider the
possibility of having a running of the
spectral index $dn_{s}/dlnk$ at $k=0.002 h^{-1}Mpc$ 
(see e.g. \cite{will}), an extra-background of
relativistic particles (parametrized with an effective number of
neutrino species $N^{\nu}_{eff}\neq3$, see e.g. \cite{bowen}) and
a non-zero, degenerate, neutrino mass of energy density
(see e.g. \cite{fogli}):
\bea
\Omega_{\nu}h^2=\frac{\Sigma m_{\nu}}{92.5 eV}
\eea
\noindent in order to establish how robust measurements of $z_{acc}$ and $z_{eq}$ are to broader cosmological models.

\noindent Finally, we will also investigate the possibility
of a dark energy equation of state $w$ different from $-1$.
Other than a constant equation of state $w$ 
we consider the possibility of a varying with redshift equation
of state. In particular we consider a
linear dependence on scale factor $a=(1+z)^{-1}$ as
\cite{Chevallier01} and \cite{Linder}:
\begin{equation}
w(a)=w_0+w_1(1-a)
\end{equation}
\noindent where the equation of state changes from 
$w_0$ to $w_0+w_1$ at higher redshifts.
We refer to this as Chevallier-Polarski-Linder (CPL) parameterization. 

We also consider a more sophisticated parametrization
that takes in to account the rate and redshift 
of the transition. We use the model proposed by Hannestad
and Mortsell (HM), see \cite{hannestad}), where:
\begin{equation}
w(a)=w_0w_1\left(\frac{a^q+a_{s}^{q}}{w_1a^q+w_0a_s^q}.\right)
\end{equation}
\noindent In this model the equation of state changes
from $w_0$ to $w_1$ around redshift $z_s=1-1/a_s$ with a gradient
transition given by $q$. We assume $w_{0,1}>-3$, 
$0.1 < a_s < 1.0$ and $1 < q < 10$
as external priors for this model.

Finally, we consider the Quartessence (or Chaplygin gas) model (see
\cite{chaplygin}) as unified dark energy-dark matter model.
In this scenario cold dark matter and dark energy are the same fluid
with equation of state:
\begin{equation}
w(a)=\frac{w_{0}}{-w_{0}+(1+w_{0}){a^{-3(\alpha +1)}}}  \label{wx}
\end{equation}%
\noindent where $w_0$ is the current value of the
equation of state and $\alpha$ is a parameter that has to be
constrained from observations.
From the equation above, it is clear that at early times, when
$a\rightarrow 0,$ we have $w\rightarrow 0$, and the fluid
behaves as non relativistic matter. At late times, when $a\gg 1$, we
obtain $w\rightarrow -1$. The matter clustering presents 
strong instabilities and oscillations in this model \cite{Sandvik:2002jz} unless one assumes 
intrinsic non-adiabatic perturbations such that the effective sound speed
vanishes \cite{Reis:2003mw}. In this paper we therefore
consider only this ``Silent'' Quartessence.

The MCMC convergence diagnostics are done on $7$ chains applying the
Gelman and Rubin ``variance of chain mean''$/$``mean of chain variances'' $R$
statistic for each parameter. Our $1-D$ and $2-D$ constraints are obtained
after marginalization over the remaining ``nuisance'' parameters, again using
the programs included in the \texttt{cosmomc} package. 
 Temperature, cross polarization and
  polarization CMB fluctuations from the WMAP 3 year data \cite{spergel,Page:2006hz,Hinshaw:2006ia,Jarosik:2006ib} are considered and we include a top-hat age prior 
$10 \mathrm{\ Gyr} <  t_0 < 20 \mathrm{\ Gyr}$.
We combine the WMAP data with the the real-space power spectrum of
galaxies from the Sloan Digital Sky Survey (SDSS) 
\cite{2004ApJ...606..702T} and 2dF survey \cite{2005MNRAS.362..505C}.
We restrict the analysis to a range of scales over which the
fluctuations are assumed to be in the linear 
regime (technically, $k < 0.2
h^{-1}$~Mpc) and we marginalize over a  bias $b$ considered as an
additional nuisance parameter. We also incorporate the constraints
obtained from the supernova (SN-Ia) luminosity measurements by using 
the so-called GOLD data set from \cite{riess} and 
 the Supernovae Legacy Survey (SNLS) data from \cite{2006A&A...447...31A}.

\section{Results}
\label{sec4}

\subsection{Cosmological Datasets}

Using the analysis method described in the previous 
section we have constrained the value of 
$z_{eq}$, $t_{eq}$, $z_{acc}$ and $t_{acc}$ in light of the various datasets and cosmological scenarios. 
This kind of test is 
extremely useful in order to identify the presence 
of possible systematics.

The constraints on $z_{eq}$, $t_{eq}$, 
$z_{acc}$ and $t_{acc}$ for various datasets 
are reported in Table I.

\begin{table}
\begin{tabular}{l|r|r|r|r|r}
\hline \hline
Dataset & $z_{eq}$& $t_0-t_{eq}$&$z_{acc}$ & $t_0-t_{acc}$ & $t_0$\\
WMAP+ & & [Gyrs]& &[Grys] & [Gyrs]\\
\hline \hline
Alone &$0.47_{-0.09}^{+0.09}$&$4.7_{-0.5}^{+0.5}$ &$0.86_{-0.12}^{+0.11}$ & $7.0_{-0.4}^{+0.4}$&$13.8_{-0.3}^{+0.3}$\\
+SDSS  &$0.40_{-0.07}^{+0.08}$&$4.3_{-0.5}^{+0.5}$& $0.77_{-0.10}^{+0.10}$ & $6.7_{-0.3}^{+0.3}$&$13.8_{-0.2}^{+0.3}$\\
+2dF  &$0.48_{-0.05}^{+0.06}$&$4.8_{-0.3}^{+0.3}$& $0.87_{-0.07}^{+0.07}$ & $7.1_{-0.3}^{+0.2}$&$13.8_{-0.2}^{+0.2}$\\
+GOLD &$0.38_{-0.06}^{+0.06}$&$4.1_{-0.4}^{+0.4}$ & $0.74_{-0.08}^{+0.08}$ & $6.6_{-0.3}^{+0.3}$&$13.8_{-0.2}^{+0.2}$\\
+SNLS &$0.45_{-0.06}^{+0.07}$&$4.6_{-0.4}^{+0.4}$ & $0.83_{-0.08}^{+0.08}$ & $6.9_{-0.3}^{+0.3}$&$13.8_{-0.1}^{+0.1}$\\
+all &$0.40_{-0.04}^{+0.04}$&$4.3_{-0.3}^{+0.3}$ & $0.76_{-0.05}^{+0.05}$ & $6.7_{-0.2}^{+0.2}$&$13.9_{-0.2}^{+0.1}$\\
\hline \hline 
\hline
\end{tabular}
\caption{Constraints on $z_{eq}$, $t_{eq}$, 
$z_{acc}$ and  $t_{acc}$, at 68\% c.l., in comparison with various datasets for $\Lambda$CDM.}
\end{table}

As we can see, there is a general agreement
between the results: namely, in a cosmological
constant model, dark energy became
the dominant component at redshift
$z_{eq} \sim 0.4$, $4.3$ Gyrs ago, and the accelerated 
expansion of the universe started at $z_{acc} \sim 0.75$,
$6.7$ Gyrs ago. Since we are assuming a cosmological
constant $z_{eq}$ and $z_{acc}$ are not independent 
but follow Eq.(3). It is 
interesting, however, to note that the SDSS
and GOLD datasets seem to favour a lower redshift for the
dark energy's dominant than that suggested when the 
2dF or SNLS datasets are included, respectively.

\subsection{Theoretical assumptions on the background cosmological model.}

Since the results appear stable to the inclusion/exclusion of the
experimental datasets, we now consider the full set of cosmological
data and study the dependence on some of the theoretical assumptions 
on the background cosmological model. We consider possible variations
from $-1$ in a constant dark energy equation of state, non-flat
universes, a running of the spectral index of the primordial 
inflationary perturbations, massive neutrinos and an extra
background of relativistic particles. All those constraints
are reported in Table II.
\begin{table}
\begin{tabular}{l|r|r|r|r|r}
\hline \hline
Model & $z_{eq}$& $t_0-t_{eq}$&$z_{acc}$ & $t_0-t_{acc}$ & $t_0$\\
 & & [Gyrs]& &[Grys] & [Gyrs]\\
\hline \hline
$w\neq -1$  & $0.48_{-0.07}^{+0.07}$ & $4.9_{-0.5}^{+0.4}$ & $0.81_{-0.06}^{+0.06}$ & $6.9_{-0.2}^{+0.2}$&$13.9_{-0.2}^{+0.1}$\\
$\Omega_{tot}\neq1$ & $0.32_{-0.10}^{+0.10}$&$3.9_{-0.8}^{+0.8}$ & $0.68_{-0.10}^{+0.10}$ & $6.9_{-0.3}^{+0.3}$&$15.1_{-0.9}^{+0.8}$\\
$dn/dlnk\neq0$ & $0.37_{-0.05}^{+0.05}$&$4.1_{-0.3}^{+0.3}$&$0.72_{-0.10}^{+0.06}$ & $6.6_{-0.2}^{+0.2}$&$14.1_{-0.2}^{+0.1}$\\
$N_{eff}^{\nu}\neq 3$ & $0.40_{-0.06}^{+0.05}$&$4.3_{-0.4}^{+0.5}$ & $0.77_{-0.06}^{+0.06}$ & $6.8_{-0.6}^{+0.6}$&$14.0_{-1.4}^{+1.2}$\\
$\Sigma m_{\nu}>0$& $0.37_{-0.04}^{+0.04}$&$4.2_{-0.3}^{+0.3}$& $0.73_{-0.05}^{+0.05}$ & $6.7_{-0.2}^{+0.2}$&$14.1_{-0.2}^{+0.2}$\\
\hline \hline 
\hline
\end{tabular}
\caption{Constraints on $z_{eq}$, $t_{eq}$, 
$z_{acc}$ and  $t_{acc}$, at 68\% c.l., under differing 
theoretical assumptions for the underlying cosmological model.}
\end{table}
As we can see, the results are consistent with those reported
in Table I. However, as expected, the constraints are in general
weaker. While including a constant dark energy equation of state 
$w\neq -1$ has little effect, considering a universe with spatial curvature
generally doubles the error bars on all the parameters and results in a lower
redshift and time of dark energy domination. It is interesting to note
that considering an extra background of relativistic particles has a
strong effect on the age of the universe and of dark energy.
This raises an interesting question of whether the recent discovery of the APM 08279+5255 quasar at $z = 3.91$, 
whose age of $2-3$ Gyr can't easily be accommodated in the standard 
scenario, could provide a hint for the presence of
an extra background of relativistic particles \cite{jain}.

\subsection{Dynamical Dark Energy}

\begin{table}
\begin{tabular}{l|r|r|r|r|r}
\hline \hline
Model & $z_{eq}$& $t_0-t_{eq}$&$z_{acc}$ & $t_0-t_{acc}$ & $t_0$\\
 & & [Gyrs]& &[Grys] & [Gyrs]\\
\hline \hline
$w\neq -1$  & $0.43_{-0.06}^{+0.07}$ & $4.5_{-0.5}^{+0.5}$ & $0.79_{-0.07}^{+0.07}$ & $6.8_{-0.3}^{+0.3}$&$13.8_{-0.2}^{+0.1}$\\
CPL  & $0.44_{-0.10}^{+0.11}$ & $4.5_{-0.6}^{+0.7}$ & $0.80_{-0.17}^{+0.16}$ & $6.8_{-0.7}^{+0.6}$&$13.9_{-0.2}^{+0.2}$\\
HM  & $0.45_{-0.10}^{+0.10}$ & $4.6_{-0.7}^{+0.6}$ & $0.79_{-0.14}^{+0.14}$ & $6.7_{-0.5}^{+0.6}$&$13.9_{-0.3}^{+0.2}$\\
SQ  & -- & -- & $0.80_{-0.08}^{+0.08}$& $6.8_{-0.3}^{+0.3}$&$13.8_{-0.2}^{+0.2}$\\
\hline \hline 
\hline
\end{tabular}
\caption{Constraints on $z_{eq}$, $t_{eq}$, 
$z_{acc}$ and  $t_{acc}$, at $68 \%$ c.l., for different
theoretical assumptions about the nature of the dark energy
component.}
\end{table}

We now study the sensitivity of the epochs of dark energy domination and the onset of acceleration to differing dark energy models. We compare models to WMAP+2dF+SNLS
datasets, this should be considered a more conservative choice in comparison to
the ``all'' dataset described in the previous section. 
For the Silent Quartessence, however, we consider only WMAP+SNLS, as a unified dark energy model, include no 
cold dark matter, we omit the redshift and time of equivalence 
with the baryonic component.
The constraints are reported in Table III.

Allowing for an equation of state which is 
varying with redshift can strongly affect the constraints,
with error bars as large as four times those reported in
Table I. However, the mean values are generally consistent 
with the previous results, i.e. there is no indication 
for deviations from a cosmological constant. 
In this sense, it is useful to plot the constraints
on the $z_{acc}-z_{eq}$ plane as we do in Fig.1. 
A cosmological constant in this plane generates a line 
described by Eq. 3.. As showed by three contour plots, 
while including a dynamical component leaves the possibility of
a different relation between the two redshifts, the case
of a cosmological constant is always well inside the $1 \sigma$ c.l..

In Fig.2 we plot the $2 \sigma$ constraints on the deceleration 
parameter $q$ in function of the redshift for the four different dark
energy parameterizations. While there is 
a large spread in the values, especially when more
complex parametrizations such as CPL or HM are considered, 
there is a very good agreement, and all the models
point towards the same acceleration redshift value at
$z \sim 0.8$.

\section{Conclusions}
\label{sec5}

\begin{figure}
\includegraphics[width=2.5in]{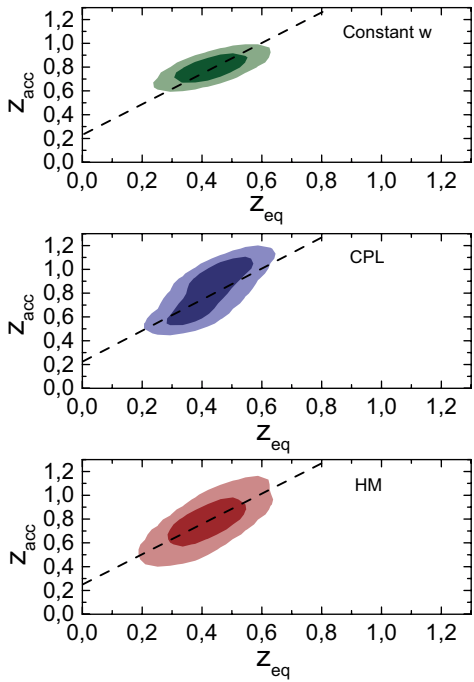}
\caption{\label{fig1} Constraints on the
$z_{acc}$---$z_{eq}$ plane for different dark energy parameterizations.
From top to bottom $w$ constant, Chevallier-Polarski-Linder (CPL)
and Hannestad Mortsell (HM) models. Also plotted (dashed line) is the
cosmological constant case $z_{acc}=2^{1/3}(z_{eq}+1)-1$}
\end{figure}

\begin{figure}
\includegraphics[width=3.in]{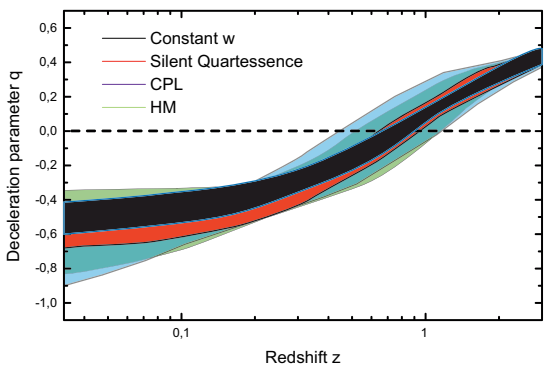}
\caption{\label{fig2} Constraints at $95 \%$ c.l. 
on the deceleration parameter $q(z)$ in function of redshift
 for different dark energy parametrizations
(see text).}
\end{figure}

In this brief paper we have presented several constraints on the
epoch and redshift of dark energy domination and of cosmic
acceleration. We have derived those constraints using different
datasets, different theoretical assumptions and different dark energy
parametrizations. We have found that a redshift and epoch of
acceleration at $z_{acc}= 0.78$ and $t_0-t_{acc}=6.9$ Gyrs 
and a redshift and epoch of dark energy domination start at
  $z_{acc}= 0.43$ and $t_0-t_{acc}=4.4$ Gyrs, as expected for
a flat universe with $\Omega_{\Lambda}=0.7$ is
 in agreement with all the possible cases considered.
Moreover, despite the large set of models and data analyzed,
there a very little spread in the best-fit values.
Curvature, running of the spectral-index, massive neutrinos and
an extra-background of relativistic particles
are non-standard cosmological parameters that can strongly enlarge
the error bars on $z_{eq}$ and $z_{acc}$ 
in case of a cosmological constant model.
Allowing a constant or dynamical dark energy equation of state 
different from $-1$ produces similar results,
however the best fit values are, again, not significantly altered.
A tension in the derived best fit values appears when considering 
galaxy clustering data from SLOAN and 2dF and supernovae type Ia
from Riess et al. and SNLS datasets separately. 
However the significance of the discrepancy is well below the 
$2 \sigma$ c.l..
As a final remark, we like to stress that the analysis presented
here relies nonetheless in the assumption of a class of scenarios.
It may be possible to construct more complicated cosmological and/or 
dark energy models that could evade the constraints presented
here.
Future data may well falsify those possibilities as 
well the simple case of a cosmological constant by
testing the $z_{acc}$---$z_{eq}$ relation of Eq. 3.

\textbf{Acknowledgments} 
We are pleased to thank Rachel Bean for help, discussions and 
useful comments on the manuscript.

\end{document}